\documentclass[journal]{IEEEtran}
\usepackage{bm}
\usepackage{cite}
\usepackage{amsmath,amssymb,amsfonts}
\usepackage{algorithm}
\usepackage{algpseudocode}
\usepackage{graphicx}
\usepackage{textcomp}
\usepackage{xcolor}
\usepackage{caption}
\usepackage{subcaption}
\allowdisplaybreaks[4]

\def\BibTeX{\rm B\kern-.05em{\sc i\kern-.025em b}\kern-.08emT\kern-.1667em\lower.7ex\hbox{E}\kern-.125emX}

\newtheorem{theorem}{Theorem}

\begin{document}

\title{Fluid Antenna Array Enhanced Over-the-Air Computation}

\author{Deyou Zhang,~\IEEEmembership{Member,~IEEE}, Sicong Ye, Ming Xiao,~\IEEEmembership{Senior Member,~IEEE}, Kezhi Wang,~\IEEEmembership{Senior Member,~IEEE}, Marco Di Renzo,~\IEEEmembership{Fellow,~IEEE}, and Mikael Skoglund,~\IEEEmembership{Fellow,~IEEE} \vspace{-4em}

\thanks{Manuscript received December 22, 2023; revised February 16, 2024; accepted March 8, 2024. This work is supported in part by Horizon Europe COVER project under grant number 101086228, and UKRI grant EP/Y028031/1. The work of M. Di Renzo is supported in part by the Horizon Europe projects COVER-101086228, UNITE-101129618, and INSTINCT-101139161, and the ANR projects NF-PERSEUS 22-PEFT-004 and PASSIONATE CHIST-ERA-22-WAI-04 (ANR-23-CHR4-0003-01). D. Zhang and S. Ye contributed equally to this work. \emph{(Corresponding author: Ming Xiao.)}}

l\thanks{D. Zhang is with the School of Electronics and Information Engineering, Beihang University, 100191 Beijing, China (email: deyou@kth.se).}
\thanks{S. Ye, M. Xiao, and M. Skoglund are with the Division of Information Science and Engineering, KTH Royal Institute of Technology, 10044 Stockholm, Sweden (email: \{sicongy, mingx, skoglund\}@kth.se).}
\thanks{K. Wang is with the Department of Computer Science, Brunel University London, Uxbridge, Middlesex, UB8 3PH (email: kezhi.wang@brunel.ac.uk).} \thanks{M. Di Renzo is with the Universit\'{e} Paris-Saclay, CNRS, CentraleSup\'{e}lec, Laboratoire des Signaux et Syst\`{e}mes, 91192 Gif-sur-Yvette, France (email: marco.di-renzo@universite-paris-saclay.fr).}
}

\maketitle

\begin{abstract}
Over-the-air computation (AirComp) has emerged as a promising technology for fast wireless data aggregation by harnessing the superposition property of wireless multiple-access channels. This paper investigates a fluid antenna (FA) array-enhanced AirComp system, employing the new degrees of freedom introduced by antenna movements. Specifically, we jointly optimize the transceiver design and antenna position vector (APV) to minimize the mean squared error (MSE) between target and estimated function values. To tackle the resulting highly non-convex problem, we adopt an alternating optimization technique to decompose it into three subproblems. These subproblems are then iteratively solved until convergence, leading to a locally optimal solution. Numerical results show that FA arrays with the proposed transceiver and APV design significantly outperform the traditional fixed-position antenna arrays in terms of MSE.
\end{abstract}


\IEEEpeerreviewmaketitle

\section{Introduction}
The rapid evolution of artificial intelligence and machine learning technologies has led to the proliferation of various intelligent services, imposing unprecedented demands for massive connectivity and fast data aggregation. However, limited radio resources and stringent latency requirements pose significant challenges in meeting these demands. In response, over-the-air computation (AirComp) has emerged \cite{GuangxuZhu-Magazine, ZhibinWang-Survey}. The fundamental principle of AirComp is to harness the superposition property of the wireless multiple-access channel to achieve over-the-air aggregation of data concurrently transmitted from multiple devices. By seamlessly integrating communication and computation procedures, AirComp enables ``compute-when-communicate'' holding the potential for ultra-fast data aggregation even across massive wireless networks.

Prior works have extensively explored AirComp, particularly from the perspective of transceiver design \cite{XiaowenCao-OTA, WanchunLiu-OTA, LiChen-SIMO-UF, Zhu-MIMO}. Specifically, authors in \cite{XiaowenCao-OTA} and \cite{WanchunLiu-OTA} concentrated on the single-input single-output (SISO) configuration, exploring optimal transceiver designs for AirComp systems. Subsequent research by \cite{LiChen-SIMO-UF} investigated AirComp in the context of a single-input multiple-output (SIMO) setup, proposing a uniform-forcing transceiver design to manage non-uniform channel fading between the access point (AP) and each edge device. To enable multi-modal sensing or multi-function computation, \cite{Zhu-MIMO} further studied a multiple-input multiple-output (MIMO) AirComp system, deriving a closed-form solution for transmit and receive beamformers design. Moreover, with the advent of reconfigurable intelligent surfaces (RISs), authors in \cite{YMShi-OTA-RIS, Zhai-STAR-RIS, My-ActiveRIS-AirComp} integrated RISs into AirComp systems, achieving substantial performance enhancements without overly complicating system complexity.

In addition to RISs, fluid antenna (FA) or movable antenna (MA) systems also emerge as a promising technology for manipulating wireless channel conditions through antenna movements, thus introducing new degrees of freedom (DoFs) \cite{KKWong-FAS-CL}. Prior works have showcased the advantages of FAs in enhancing multi-beamforming \cite{RuiZhang-MA-BF}, achieving spatial diversity gain \cite{KKWong-FAS-TWC}, and minimizing total transmit power \cite{Derrick-MA} compared to fixed-position antennas (FPAs). However, to the best of our knowledge, the integration of FAs into AirComp systems remains unexplored in current literature.

This paper addresses this gap by investigating an FA array-enhanced AirComp system, wherein an AP equipped with an FA array aggregates data concurrently transmitted from multiple users via AirComp. Our main contributions are summarized as follows:

1) We seek to minimize the mean squared error (MSE) between target and estimated function values by jointly optimizing the transceiver design and antenna position vector (APV). To handle the resulting highly non-convex problem, we adopt an alternating optimization (AO) technique to decompose it into three subproblems, corresponding to $a)$ the transmit equalization coefficients of users, $b)$ the AP decoding vector, and $c)$ the APV.

2) The first two subproblems are convex quadratically constrained quadratic programs (QCQPs) and can be efficiently solved using off-the-shelf solvers. However, optimizing the APV remains a non-convex problem, for which we employ the primal-dual interior point (PDIP) method. By iteratively solving these subproblems until convergence, we achieve a locally optimal solution for the original problem.

3) Numerical results demonstrate that FA arrays with the proposed transceiver and APV design significantly outperform traditional FPAs in terms of MSE.

Throughout this paper, we use regular, bold lowercase, and bold uppercase letters to denote scalars, vectors, and matrices, respectively; $\mathcal R$ and $\mathcal C$ to denote real and complex number sets, respectively; $(\cdot)^T$ and $(\cdot)^H$ to denote the transpose and conjugate transpose, respectively. We use $x_i$ or $[\bm x]_i$ to denote the $i$-th entry in $\bm x$; $\|\bm x\|$ to denote the $\ell_2$-norm of $\bm x$; ${\rm diag}(\bm x)$ to denote a diagonal matrix with its diagonal entries specified by $\bm x$. We use $\bm I$ to denote the identity matrix, $\nabla$ to denote the gradient operator, and ${\mathbb E}$ to denote the expectation operator.

\section{System Model and Problem Formulation}\label{Sec-SM}
\subsection{System Model}
\begin{figure}
\centering
\includegraphics[width = 7.6cm]{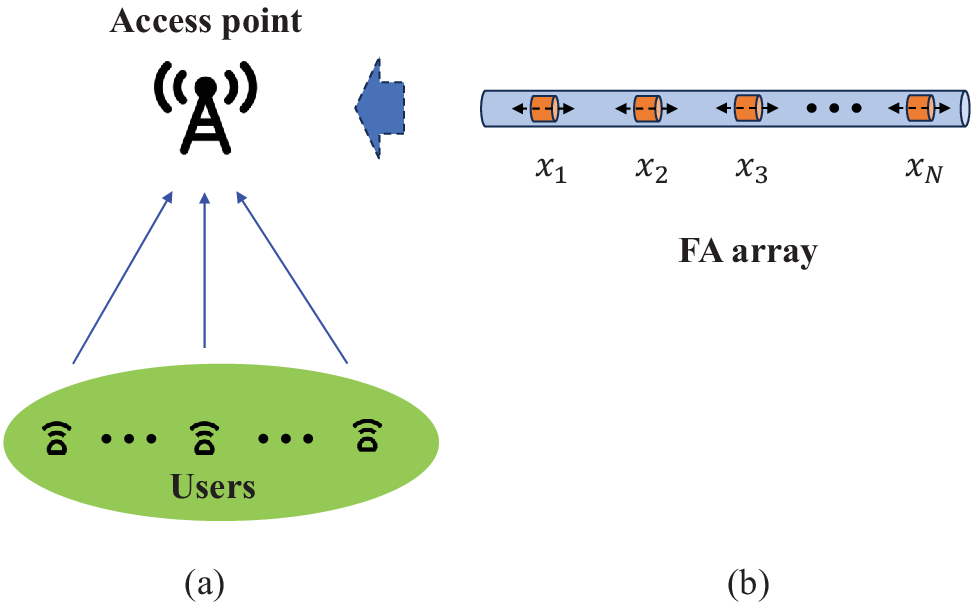}
\caption{Illustration of the considered system model.}\label{Fig-SM}
\vspace{-1em}
\end{figure}

Consider a multi-user SIMO communication system consisting of $K$ single-antenna users and an AP with $N$ antennas, as shown in Fig. \ref{Fig-SM}(a). Let $s_k \in {\mathcal C}$ represent the data generated by user $k$, $\forall k \in {\cal K} \triangleq \{1, \cdots, K\}$. For simplicity, we assume that $s_k$ possesses zero mean and unit power, i.e., ${\mathbb E}[s_k] = 0$, and ${\mathbb E}[|s_k|^2] = 1$, $\forall k \in {\cal K}$, and that $s_1, \cdots, s_K$ are uncorrelated, i.e., ${\mathbb E}[s_k s^*_j] = 0$, $\forall k \ne j$. The objective of this paper is to recover the summation of all users' data
\begin{equation}
    s = \sum\limits_{k=1}^K s_k.
\end{equation}
at the AP by harnessing the superposition property of the wireless multiple-access channel.

As shown in Fig. \ref{Fig-SM}(b), the AP is assumed to be equipped with an FA array, where the positions of the $N$ FAs can be adjusted within a one-dimensional line segment of length $L$. Let $x_n \in [0, L]$ denote the position of the $n$-th FA, and $\bm x = [x_1, \cdots, x_N]^T$ the APV of all $N$ FAs, satisfying $0 \le x_1 < \cdots < x_N \le L$ without loss of generality. Consequently, the receive steering vector of the FA array can be expressed as a function of the APV $\bm x$ and the steering angle $\theta$:
\begin{equation}
    \bm a(\bm x, \theta) = \left[e^{j \frac{2\pi}{\lambda} x_1 \cos(\theta)}, \cdots, e^{j \frac{2\pi}{\lambda} x_N \cos(\theta)}\right]^T,
\end{equation}
where $\lambda$ denotes the wavelength.

Let $\bm h_k \in {\cal C}^{N \times 1}$ denote the channel from user $k$ to AP, given by\footnote{As in \cite{RuiZhang-MA-BF}, since the LoS path is usually much stronger than the non-LoS paths, we thus ignore the non-LoS paths for simplicity.}
\begin{equation}
    \bm h_k = \alpha_k \bm a(\bm x, \theta_k),
\end{equation}
where $\alpha_k$ and $\theta_k$ represent the propagation gain and angle of arrival (AoA) of the line-of-sight (LoS) path, respectively. Given $\{\bm h_k\}$, we can express the received signal at AP as follows:
\begin{equation}
    \bm y = \sum\limits_{k=1}^K \bm h_k b_k s_k + \bm z,
\end{equation}
where $b_k$ is the transmit equalization coefficient of user $k$, $\forall k \in \mathcal K$, and $\bm z \sim {\cal CN}(\bm 0, \sigma^2 \bm I)$ is the additive white Gaussian noise at AP.

With a decoding vector $\bm m \in {\cal C}^{N \times 1}$ at AP, the estimated target function is given by
\begin{equation}
    \hat s = \bm m^H \bm y = \sum\limits_{k=1}^K \bm m^H \bm h_k b_k s_k + \bm m^H \bm z.
\end{equation}

\subsection{Problem Formulation}
In this paper, we aim to minimize the distortion between the target and the estimated function variables, which is measured by the MSE defined as follows:
\begin{align}
    \mathbb{MSE} & \triangleq {\mathbb E}[|\hat s - s|^2] \nonumber \\[1ex]
    & = {\mathbb E}\left[\left|\sum\limits_{k=1}^K \left(\bm m^H \bm h_k b_k - 1\right) s_k + \bm m^H \bm z\right|^2\right] \nonumber \\[1ex]
    & = \sum\limits_{k=1}^K \left|\bm m^H \bm h_k b_k - 1 \right|^2 + \sigma^2 \|\bm m\|^2.
\end{align}
To minimize the MSE, we need to seek the optimal $\bm b \triangleq [b_1, \cdots, b_K]^T$, $\bm m$, and $\bm x$, leading to the following optimization problem:
\begin{subequations}\label{OP0}
    \begin{align}
        \min\limits_{\bm b, \bm m, \bm x} &~\sum\limits_{k=1}^K \left|\bm m^H \bm h_k (\bm x) b_k - 1 \right|^2 + \sigma^2 \|\bm m\|^2 \\[1ex]
        \text{s.t.}~&~~ |b_k|^2 \le P_k, ~\forall k \in \mathcal K, \label{OP0-UserPowerCons} \\[1ex]
        &~~ x_1 \ge 0, x_N \le L, \label{OP0-APVCons1} \\[1ex]
        &~~ x_n - x_{n-1} \ge L_0, ~\forall n = 2, \cdots, N, \label{OP0-APVCons2}
    \end{align}
\end{subequations}
where \eqref{OP0-UserPowerCons} accounts for the power constraint for each user, with $P_k$ denoting the maximum transmission power of user $k$, $\forall k \in \mathcal K$, \eqref{OP0-APVCons1} guarantees that the FAs are moved within the feasible region $[0, L]$, and \eqref{OP0-APVCons2} ensures that the distance between two adjacent FAs is no less than $L_0$ to avoid antenna coupling.

\section{Transceiver and APV Design}
In the sequel, we adopt the AO technique to resolve the coupling between $\bm b$, $\bm m$, and $\bm x$ in \eqref{OP0}, and optimize one variable at a time with others being fixed.

1) Optimization of $\bm b$: The associated optimization problem with respect to $\bm b$ is given by
\begin{subequations}\label{Opt-b}
    \begin{align}
        \min\limits_{\bm b} &~\sum\limits_{k=1}^K \left|\bm m^H \bm h_k b_k - 1 \right|^2 \\[1ex]
        \text{s.t.} &~~ |b_k|^2 \le P_k, ~\forall k \in \mathcal K.
    \end{align}
\end{subequations}
It is observed that $b_1, \cdots, b_K$ in \eqref{Opt-b} are decoupled and can be decomposed into $K$ subproblems. The subproblem associated with $b_k$, $\forall k \in \mathcal K$, is given by
\begin{subequations}\label{Opt-b2}
    \begin{align}
        \min\limits_{b_k} &~\left|\bm m^H \bm h_k b_k - 1 \right|^2 \\[1ex]
        \text{s.t.} &~~ |b_k|^2 \le P_k,
    \end{align}
\end{subequations}
which is a convex QCQP and can be solved optimally using the Karush-Kuhn-Tucker (KKT) conditions, as detailed below.

Firstly, the Lagrangian associated with \eqref{Opt-b2} can be expressed as follows:
\begin{equation}\label{Opt-b2-Lagrangian}
    {\mathcal L}_1(b_k, \mu_k) = \left|\bm m^H \bm h_k b_k - 1 \right|^2 + \mu_k \left(|b_k|^2 - P_k\right),
\end{equation}
where $\mu_k \ge 0$ is the Lagrange multiplier. The KKT conditions of \eqref{Opt-b2-Lagrangian} are given by
\begin{subequations}
    \begin{align}
        \frac{\partial \mathcal L_1}{\partial b^*_k} = |\bm m^H \bm h_k|^2 b_k - \bm h^H_k \bm m + \mu_k b_k = 0, \label{Opt-b2-KKT-b} \\[1ex]
        \mu_k \left(|b_k|^2 - P_k\right) = 0, \label{Opt-b2-KKT-Slackness} \\[1ex]
        |b_k|^2 \le P_k. \label{Opt-b2-KKT-PowerCons}
     \end{align}
\end{subequations}
From \eqref{Opt-b2-KKT-b}, we can derive the optimal $b_k$ as follows:
\begin{equation}\label{Optimal-b}
    b_k^{\star} = \frac{\bm h^H_k \bm m}{|\bm m^H \bm h_k|^2 + \mu_k},
\end{equation}
where the nonnegative Lagrange multiplier $\mu_k$ should be chosen to satisfy \eqref{Opt-b2-KKT-Slackness} and \eqref{Opt-b2-KKT-PowerCons}. By substituting \eqref{Optimal-b} into \eqref{Opt-b2-KKT-Slackness}, we obtain that
\begin{equation}
    \mu_k^{\star} = \max\left(\frac{|\bm m^H \bm h_k|}{\sqrt{P_k}} - |\bm m^H \bm h_k|^2, 0\right).
\end{equation}

2) Optimization of $\bm m$: The associated optimization problem with respect to $\bm m$ is given by
\begin{equation}\label{Opt-m}
    \min\limits_{\bm m}~\mathbb{MSE}(\bm m) = \sum\limits_{k=1}^K \left|\bm m^H \bm h_k b_k - 1 \right|^2 + \sigma^2 \|\bm m\|^2.
\end{equation}
It is observed that \eqref{Opt-m} is a (convex) least squares problem, and the optimal $\bm m$ can be found by setting the derivative $\mathbb{MSE}(\bm m) / \partial \bm m^*$ to zero. Specifically,
\begin{equation}
    \frac{\mathbb{MSE}(\bm m)}{\partial \bm m^*} = \sum\limits_{k=1}^K \left(|b_k|^2 \bm h_k \bm h_k^H \bm m - b_k \bm h_k\right) + \sigma^2 \bm m = \bm 0,
\end{equation}
which yields
\begin{equation}\label{Optimal-m}
    \bm m^{\star} = \left(\sigma^2 \bm I + \sum\limits_{k=1}^K |b_k|^2 \bm h_k \bm h_k^H \right)^{-1} \sum\limits_{k=1}^K b_k \bm h_k.
\end{equation}

3) Optimization of $\bm x$: The associated optimization problem with respect to $\bm x$ is given by
\begin{subequations}\label{Opt-x}
    \begin{align}
        \min\limits_{\bm x} &~\sum\limits_{k=1}^K \left|\bm m^H \bm a(\bm x, \theta_k) \alpha_k b_k - 1 \right|^2 \label{Opt-x-obj} \\[1ex]
        \text{s.t.} &~~\eqref{OP0-APVCons1},~\eqref{OP0-APVCons2}.
    \end{align}
\end{subequations}
Before solving \eqref{Opt-x}, we rewrite \eqref{Opt-x-obj} into a more tractable form. Firstly, we expand $\sum\nolimits_{k=1}^K \left|\bm m^H \bm a(\bm x, \theta_k) \alpha_k b_k - 1 \right|^2$ as follows:
\begin{equation*}
    \sum\limits_{k=1}^K \Big(|\bm w_k^H \bm a(\bm x, \theta_k)|^2 - \bm w_k^H \bm a(\bm x, \theta_k) - \bm a^H(\bm x, \theta_k) \bm w_k + 1 \Big), \nonumber
\end{equation*}
where $\bm w^H_k \triangleq \alpha_k b_k \bm m^H$, $\forall k \in \mathcal K$. By denoting the $n$-th entry in $\bm w_k$ as $w_{k,n} = |w_{k,n}| e^{j \angle w_{k,n}}$, $\forall n \in \mathcal N$, we can then rewrite $\bm w_k^H \bm a(\bm x, \theta_k)$ as follows:
\begin{align}\label{Eq-WAProduct}
    \bm w_k^H \bm a(\bm x, \theta_k) & = \sum\limits_{n=1}^N w^*_{k,n} e^{j \phi_k x_n} \nonumber \\[1ex]  
    & = \sum\limits_{n=1}^N |w_{k,n}| e^{j (\phi_k x_n - \angle w_{k,n})},
\end{align}
where $\phi_k = \frac{2 \pi}{\lambda} \cos(\theta_k)$. Given \eqref{Eq-WAProduct}, we can rewrite $F_{k}(\bm x) \triangleq |\bm w_k^H \bm a(\bm x, \theta_k)|^2$ and $G_{k}(\bm x) \triangleq \bm w_k^H \bm a(\bm x, \theta_k) + \bm a^H(\bm x, \theta_k) \bm w_k$ as follows:  
\begin{align}
    F_{k}(\bm x) & \triangleq |\bm w_k^H \bm a(\bm x, \theta_k)|^2 = \left|\sum\limits_{n=1}^N |w_{k,n}| e^{j (\phi_k x_n - \angle w_{k,n})}\right|^2 \nonumber \\[1ex]
    & = \sum\limits_{n=1}^N \sum\limits_{l =1}^N |w_{k,n}| |w_{k,l}| e^{j [\phi_k (x_n - x_l) - (\angle w_{k,n} - \angle w_{k,l})]} \nonumber \\[1ex]
    & = \sum\limits_{n=1}^N \sum\limits_{l = 1}^N |w_{k,n}| |w_{k,l}| \cos[q_k(x_n, x_l)], \label{Eq-Fk} \\[1ex]
    G_{k}(\bm x) & \triangleq \bm w_k^H \bm a(\bm x, \theta_k) + \bm a^H(\bm x, \theta_k) \bm w_k \nonumber \\[1ex]
    & = 2 \sum\limits_{n=1}^N |w_{k,n}| \cos(\phi_k x_n - \angle w_{k,n}), \label{Eq-Gk}
\end{align}
where $q_k(x_n, x_l) = \phi_k (x_n - x_l) - (\angle w_{k,n} - \angle w_{k,l})$.

Given \eqref{Eq-Fk} and \eqref{Eq-Gk}, we successfully transform \eqref{Opt-x-obj} into a more tractable form. However, the function $\sum\nolimits_{k=1}^K [F_k(\bm x) - G_k(\bm x)]$ is still highly non-convex, and thus we employ the PDIP method to obtain a locally optimal $\bm x$, as detailed in Appendix \ref{PDIP}.

The proposed transceiver and APV design approach is outlined in \textbf{Algorithm \ref{Alg-AO}}. Moreover, the convergence of \textbf{Algorithm \ref{Alg-AO}} is illustrated in the following theorem.

\begin{theorem}
    \textbf{Algorithm \ref{Alg-AO}} converges to a locally optimal point of problem \eqref{OP0} after several iterations.
\end{theorem}
\begin{IEEEproof}
    To prove the convergence of \textbf{Algorithm \ref{Alg-AO}}, we introduce superscript $t$ as the iteration index, e.g., $\bm m^t$ represents the decoding vector at the end of the $t$-th iteration round. Then, \textbf{Algorithm \ref{Alg-AO}} converges as
    \begin{align*}
        \mathbb{MSE}(\bm m^t, \bm b^t, \bm x^t) & \overset{(a)}{\ge} \mathbb{MSE}( \bm m^{t+1}, \bm b^t, \bm x^t) \\[1ex]
        & \overset{(b)}{\ge} \mathbb{MSE}(\bm m^{t+1}, \bm b^{t+1}, \bm x^t) \\[1ex]
        & \overset{(c)}{\ge} \mathbb{MSE}(\bm m^{t+1}, \bm b^{t+1}, \bm x^{t+1}),
    \end{align*}
    where $(a)$, $(b)$, and $(c)$ follow since the updates of $\bm m$, $\bm b$, and $\bm \Phi$ are the optimal (or locally optimal) solutions to \eqref{Opt-m}, \eqref{Opt-b}, and \eqref{Opt-x}, respectively. Since $\mathbb{MSE}(\bm m^t, \bm b^t, \bm x^t)$ is monotonically non-increasing in each iteration and its value is lower bounded by zero, we prove that \textbf{Algorithm \ref{Alg-AO}} will converge to a local optimum of problem \eqref{OP0} after several iterations.
\end{IEEEproof}

\begin{algorithm}[!htbp]
    \caption{Pseudo-Code for the Proposed Transceiver and APV Design}
    \begin{algorithmic}[1]
        \State Given $b_k = \sqrt{P_k}$, $\forall k \in \mathcal K$, and $x_n = \frac{Ln}{N-1}$, $\forall n \in \{0, \cdots, N-1\}$;
        \While{not converge}
        \State Update $\bm m$ through \eqref{Optimal-m};
        \State Update $\bm b$ through solving \eqref{Optimal-b};
        \State Update $\bm x$ via the PDIP method.
        \EndWhile
    \end{algorithmic}\label{Alg-AO}
\end{algorithm}

\section{Numerical Results}
This section presents numerical results demonstrating the effectiveness of introducing FA arrays in enhancing AirComp performance. We set $L_0 = 0.5 \lambda$, $L = N \lambda$, $P_k = P_0$, $\forall k \in \mathcal K$, and $\text{SNR} \triangleq P_0 / \sigma^2$. Moreover, the following three benchmark schemes are considered for comparison.
\begin{itemize}
    \item \textbf{Fixed-Position Antenna Array (FPA)}: The $N$ FAs are fixed and uniformly distributed in the feasible region $[0, L]$, i.e., the APV is fixed to $\bm x = \frac{L}{N-1} \times \left[0, 1, \cdots, N-1\right]^T$. The optimal $\bm b$ and $\bm m$ are obtained by recursively running \eqref{Optimal-b} and \eqref{Optimal-m} until convergence.
    \item \textbf{SCA}: We optimize $\bm x$ by solving \eqref{Opt-x} via the successive convex approximation (SCA) technique \cite{RuiZhang-MA-BF}, while the optimization procedures for $\bm b$ and $\bm m$ remain unchanged. For more details, refer to Appendix \ref{SCA}.
    \item \textbf{PGD}: Problem \eqref{Opt-x} for $\bm x$ can also be solved by using the projected gradient descent (PGD) algorithm \cite{PGD}, while the optimization procedures for $\bm b$ and $\bm m$ remain unchanged.
\end{itemize}

\begin{figure}
    \vskip-4pt
    \centering
    \begin{subfigure}{0.24\textwidth}
        \centering
        \includegraphics[width = \textwidth]{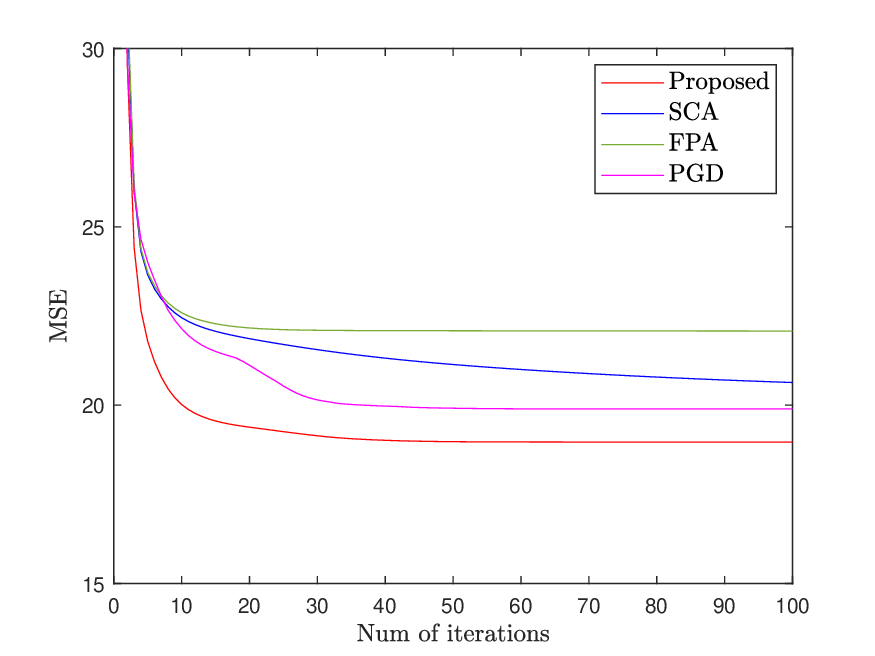}
        \caption{MSE versus the number of iterations, where $\text{SNR} = -10$ dB, $N = 10$, and $K = 100$.}\label{Fig-MSE-ITER}
    \end{subfigure}
    \begin{subfigure}{0.24\textwidth}
        \centering
        \includegraphics[width = \textwidth]{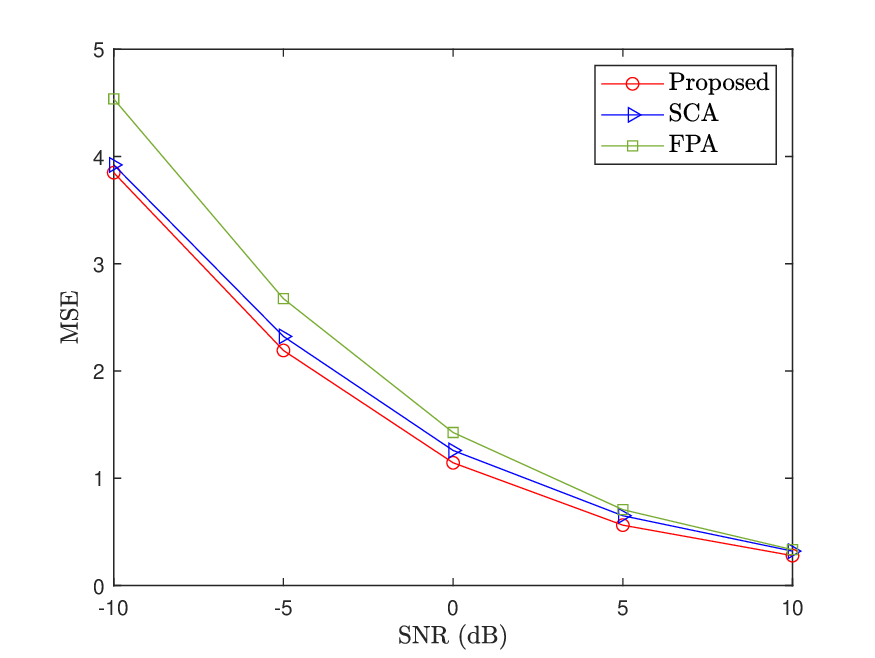}
        \caption{MSE versus SNR, where $N = 10$, and $K = 10$.}\label{Fig-MSE-SNR}
        \vspace{1em}
    \end{subfigure}
    \begin{subfigure}{0.24\textwidth}
        \centering
        \includegraphics[width = \textwidth]{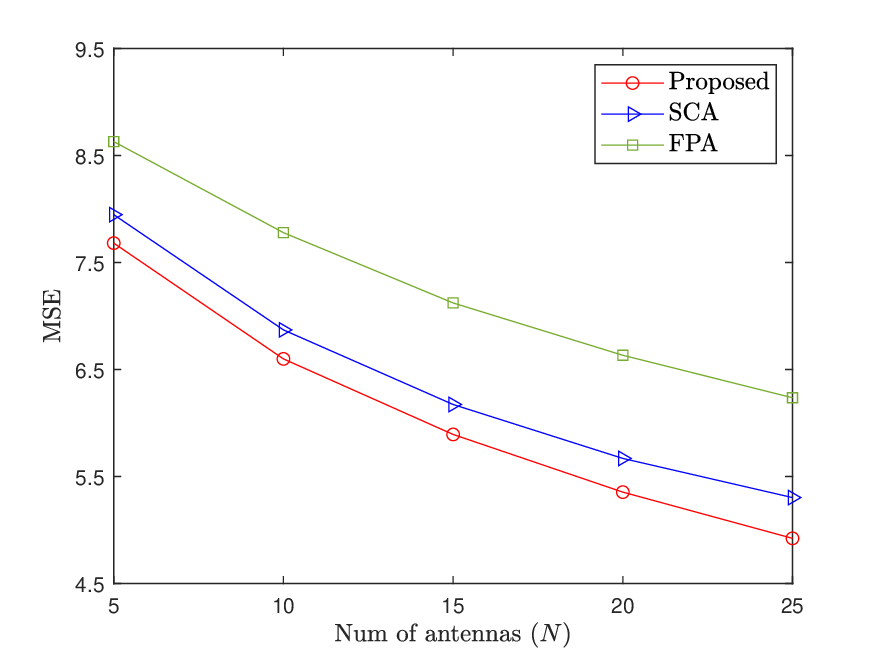}
        \caption{MSE versus the number of FAs $N$, where $\text{SNR} = -10$ dB, and $K = 10$.}\label{Fig-MSE-N}
    \end{subfigure}
    \begin{subfigure}{0.24\textwidth}
        \centering
        \includegraphics[width = \textwidth]{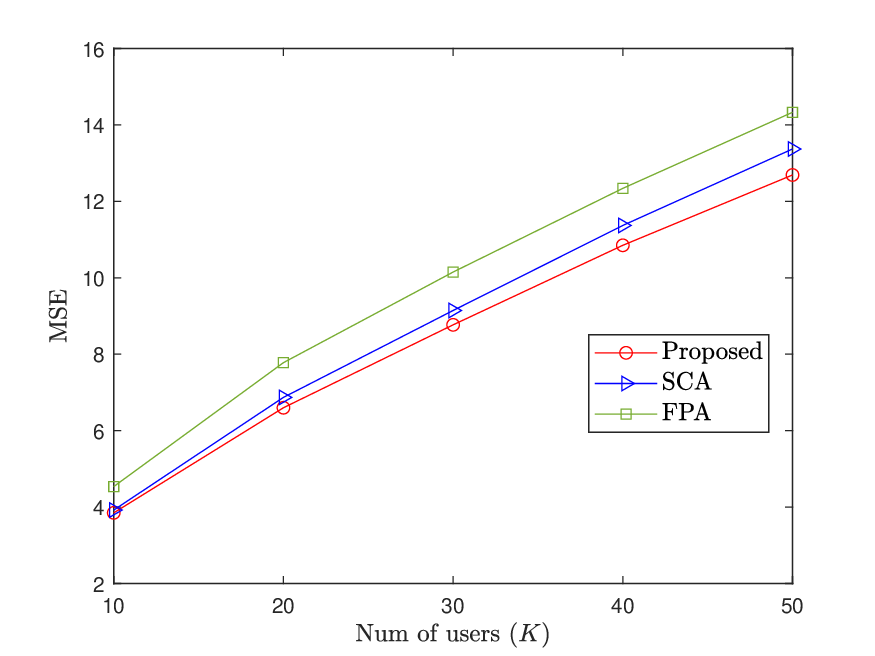}
        \caption{MSE versus the number of users, where $\text{SNR} = -10$ dB and $N = 10$.}\label{Fig-MSE-User}
    \end{subfigure}
    \caption{MSE performance versus the number of iterations, SNR, the number of FAs, and the number of users}
\end{figure}

To show the convergence of \textbf{Algorithm \ref{Alg-AO}}, we present the MSE performance across iterations in Fig. \ref{Fig-MSE-ITER}. From this figure, we observe that \textbf{Algorithm \ref{Alg-AO}} converges after a few dozens of iterations. In contrast, the SCA scheme takes $100$ iterations to converge. Additionally, we observe that \textbf{Algorithm \ref{Alg-AO}} is superior to the three benchmark schemes in terms of the MSE performance.

In Fig. \ref{Fig-MSE-SNR}, we show the MSE performance across different SNR levels. From this figure, it is observed that our proposed scheme (i.e., \textbf{Algorithm \ref{Alg-AO}}) outperforms the FPA scheme for all SNR values, though the performance gap becomes narrower as the SNR increases.

Fig. \ref{Fig-MSE-N} depicts the MSE performance across different numbers of FAs, i.e., $N$. As expected, the MSE performance consistently decreases as $N$ increases from $5$ to $25$. It is also observed that our proposed scheme showcases a significant advantage over the FPA scheme, again highlighting the benefits of optimizing the positions of FAs.

Fig. \ref{Fig-MSE-User} depicts the MSE performance across different user numbers, i.e., $K$. Unsurprisingly, the MSE performance consistently rises as $K$ increases. Notably, we also observe the consistent outperformance of the proposed scheme compared to the SCA and FPA schemes, with the performance gap widening as $K$ increases.

\section{Conclusions}\label{Section-CN}
This paper studied an FA array-enhanced AirComp system. Compared to traditional FPAs, FAs provide new DoFs through antenna movements, thus offering the potential for improved performance. As such, we jointly optimized the transceiver design and APV to minimize the MSE between target and estimated function values. Numerical results demonstrated that FA arrays with our proposed transceiver and APV design significantly outperformed traditional FPAs in terms of MSE.

Thanks to the benefits of reducing AirComp MSE, FA arrays can also be used to improve the performance of wireless federated learning systems that rely on AirComp, which is an interesting topic for future work.

\begin{appendices}
\section{}\label{PDIP}
For ease of exposition, we rewrite $x_1 \ge 0$, $x_n \le L$, and \eqref{OP0-APVCons2} as follows:
\begin{align*}
    & f_1(\bm x) = - \bm e^T_1 \bm x \le 0, \\
    & f_2(\bm x) = \bm e^T_N \bm x - L \le 0, \\
    & f_3(\bm x) = (\bm e_1 - \bm e_2)^T \bm x + L_0 \le 0, \\
    &~~\vdots \\
    & f_{N+1}(\bm x) = (\bm e_{N-1} - \bm e_N)^T \bm x + L_0 \le 0,
\end{align*}
where $\bm e_n$ denotes the $n$-th column of $\bm I_N$, $\forall n \in \{1, \cdots, N\}$. Given $g(\bm x) \triangleq \sum\nolimits_{k=1}^K [F_k(\bm x) - G_k(\bm x)]$ and $f_i(\bm x)$, $\forall i = 1,\cdots,N+1$, we rewrite \eqref{Opt-x} as follows:
\begin{subequations}\label{Opt-x2}
    \begin{align}
        \min\limits_{\bm x} &~g(\bm x) \\[1ex]
        \text{s.t.}&~f_i(\bm x) \le 0,~\forall i = 1,\cdots,N+1.
    \end{align}
\end{subequations}
The Lagrangian associated with \eqref{Opt-x2} is given by
\begin{equation}\label{Opt-x2-Lagrangian}
    \mathcal L_2(\bm x, \bm \nu) = g(\bm x) + \bm f(\bm x)^T \bm \nu,
\end{equation}
where $\bm \nu = [\nu_1, \cdots, \nu_{N+1}]^T$ is the Lagrange multiplier and $\bm f(\bm x) \triangleq [f_1(\bm x), \cdots, f_{N+1}(\bm x)]^T$. According to \cite{ConvexOpti}, the dual residual and the centrality residual corresponding to \eqref{Opt-x2-Lagrangian} can be expressed as follows:
\begin{align}
    & \bm r_{\rm dual}(\bm x, \bm \nu) \triangleq \nabla g(\bm x) + D \bm f(\bm x)^T \bm \nu, \\[1ex]
    & \bm r_{\rm cent}(\bm x, \bm \nu) \triangleq -{\rm diag}(\bm \nu) \bm f(\bm x) - (1/\delta) \bm 1,
 \end{align}
where $D \bm f(\bm x)^T = [\nabla f_1(\bm x), \cdots, \nabla f_{N+1}(\bm x)]$, and $\delta > 0$. If the current point $\bm x$ and $\bm \nu$ satisfy $\bm r_{\rm dual}(\bm x, \bm \nu) = \bm 0$, and $\bm r_{\rm cent}(\bm x, \bm \nu) = \bm 0$, then $\bm x$ is primal feasible, and $\bm \nu$ is dual feasible, with duality gap no less than $(N+1)/\delta$. Otherwise, we have to proceed with the iteration, and the primal-dual search direction $(\Delta \bm x_{\rm pd}, \Delta \bm \nu_{\rm pd})$ is determined by the solution of the following equation:
\begin{align}\label{Eq-PDSD}
\hspace{-0.15cm}\begin{bmatrix}
    \nabla^2 g(\bm x) + \sum\nolimits_{i=1}^{N+1} \nu_i \nabla^2 f_i(\bm x) & D \bm f(\bm x)^T \\[1ex]
    -{\rm diag}(\bm \nu) D \bm f(\bm x) & -{\rm diag}(\bm f(\bm x))
\end{bmatrix}
\begin{bmatrix}
    \Delta \bm x \\[1ex]
    \Delta \bm \nu
\end{bmatrix}~
\nonumber \\[1ex]
= -
\begin{bmatrix}
    \bm r_{\rm dual}(\bm x, \bm \nu) \\[1ex]
    \bm r_{\rm cent}(\bm x, \bm \nu)
\end{bmatrix}
.
\end{align}
The overall procedures of the PDIP method are summarized in \textbf{Algorithm \ref{Alg-PDIP}}.
\begin{algorithm}[!htbp]
    \caption{Pseudo-Code for Solving \eqref{Opt-x2} with the PDIP Method}
    \begin{algorithmic}[1]
        \State Choose $\bm x$ that satisfies $\bm f(\bm x) \prec \bm 0$, $\bm \nu \succ \bm 0$, $\eta \triangleq -\bm f(\bm x)^T \bm \nu$, $\epsilon > 0$, $\epsilon_{\rm feas} > 0$, and $\xi > 1$.
        \While{$\|\bm r_{\rm dual}\| > \epsilon_{\rm feas}$ or $\eta > \epsilon$}
        \State Define $\delta = \xi (N+1) / \eta$;
        \State Compute $(\Delta \bm x_{\rm pd}, \Delta \bm \nu_{\rm pd})$ via solving \eqref{Eq-PDSD};
        \State Determine step size $\gamma$ via backtracking line search;
        \State Update $(\bm x, \bm \nu) \leftarrow (\bm x, \bm \nu) + \gamma (\Delta \bm x_{\rm pd}, \Delta \bm \nu_{\rm pd})$;
        \State Compute $\eta = -\bm f(\bm x)^T \bm \nu$.
        \EndWhile
    \end{algorithmic}\label{Alg-PDIP}
\end{algorithm}

\section{}\label{SCA}
Motivated by \cite{RuiZhang-MA-BF}, we construct convex surrogate functions to locally approximate $F_{k}(\bm x)$ and $G_{k}(\bm x)$ based on the second-order Taylor expansion. Specifically, for a given $u_0 \in \mathcal R$, the second-order Taylor expansion of $\cos(u)$ is given by
\begin{equation}\label{cos-approx}
    \cos(u) \approx \cos(u_0) - \sin(u_0) (u - u_0) - \frac{1}{2} \cos(u_0) (u - u_0)^2
\end{equation}
Since $-1 \le \cos(u_0) \le 1$, $\forall u_0 \in \mathcal R$, we have
\begin{align}
    \cos(u) \overset{(a)}{\le} \cos(u_0) - \sin(u_0) (u - u_0) + \frac{1}{2} (u - u_0)^2, \label{cos-ub}
    \\
    \cos(u) \overset{(b)}{\ge} \cos(u_0) - \sin(u_0) (u - u_0) - \frac{1}{2} (u - u_0)^2, \label{cos-lb}
\end{align}
where $(a)$ is due to $\cos(u_0) \ge -1$, and $(b)$ is due to $\cos(u_0) \le 1$. Denote the $i$-th iteration of SCA as $\bm x^i = [x^i_1, \cdots, x^i_N]^T$. Given \eqref{cos-ub} and letting $u \leftarrow \phi_k (x_n - x_l) - (\angle w_{k,n} - \angle w_{k,l})$, $u_0 \leftarrow \phi_k (x^i_n - x^i_l) - (\angle w_{k,n} - \angle w_{k,l})$, we derive an upper bound for $F_k(\bm x)$, given by
\begin{equation}\label{Fk-UB}
    F_k(\bm x) \le F_k(\bm x | \bm x^i) = \bm x^T \bm A_k \bm x - (\bm v^i_k)^T \bm x + C^i_k.
\end{equation}
In \eqref{Fk-UB}, $\bm A_k$, $\bm v^i_k$, and $C^i_k$ are respectively given by
\begin{align*}
    & \bm A_k = \phi_k^2 \left(\bar w_{k,0} {\rm diag}(\bar{\bm w}_k) - \bar{\bm w}_k \bar{\bm w}_k^T\right), \\
    & [\bm v^i_k]_n = \sum\limits_{l=1}^N \left(\varphi^i_{k,n,l} - \varphi^i_{k,l,n}\right), \\
    & \varphi^i_{k,n,l} = {\bar w}_{k,n} {\bar w}_{k,l} [\sin(q_k(x_n^i, x_l^i)) \phi_k + \phi^2_k (x^i_n - x^i_l)], \\
    & C^i_k = \sum\limits_{n=1}^N \sum\limits_{l=1}^N {\bar w}_{k,n} {\bar w}_{k,l} [0.5 \phi^2_k (x^i_n - x^i_l)^2 \\
    &~~~~~~~ + \cos(q_k(x_n^i, x_l^i)) + \sin(q_k(x_n^i, x_l^i)) \phi_k (x^i_n - x^i_l)],
\end{align*}
where ${\bar w}_{k, n} = |w_{k,n}|$, $\forall k, n$, $\bar {\bm w}_k = [{\bar w}_{k, 1}, \cdots, {\bar w}_{k, N}]^T$, $\bar w_{k,0} = {\bar w}_{k, 1} + \cdots + {\bar w}_{k, N}$, and $q_k(x_n^i, x_l^i) = \phi_k (x^i_n - x^i_l) - (\angle w_{k,n} - \angle w_{k,l})$.

On the other hand, given \eqref{cos-lb} and letting $u \leftarrow \phi_k (x_n - x_l) - (\angle w_{k,n} - \angle w_{k,l})$, $u_0 \leftarrow \phi_k (x^i_n - x^i_l) - (\angle w_{k,n} - \angle w_{k,l})$, we derive a lower bound for $G_k(\bm x)$, given by
\begin{equation}\label{Gk-LB}
    G_k(\bm x) \ge G_k(\bm x | \bm x^i) = -\bm x^T \tilde{\bm A}_k \bm x + 2 (\tilde{\bm v}^i_k)^T \bm x + 2 \tilde C^i_k,
\end{equation}
where $\tilde{\bm A}_k$, $\tilde{\bm v}^i_k$, and $\tilde C^i_k$ are respectively given by
\begin{align*}
    & \tilde{\bm A}_k = \phi_k^2 {\rm diag}(\bm{\bar w}_k), \\
    & [\tilde{\bm v}^i_k]_n = \bar{w}_{k,n} [\phi^2_k x^i_n -\sin(\phi_k x_n^i - \angle w_{k,n}) \phi_k], \\
    &\tilde C^i_k = \sum\limits_{n=1}^N \bar{w}_{k,n} [\cos(\phi_k x_n^i - \angle w_{k,n}) \\
    &~~~~ + \sin(\phi_k x_n^i - \angle w_{k,n}) \phi_k x^i_n - 0.5 \phi^2_k (x^i_n)^2].
\end{align*}
Given \eqref{Fk-UB} and \eqref{Gk-LB}, the $(i+1)$-th iteration of SCA can be formulated as follows:
\begin{subequations}\label{Opt-x3}
    \begin{align}
        \min\limits_{\bm x} &~\bm x^T \bm A \bm x - (\bm v^i)^T \bm x + C^i \label{Opt-x3-obj} \\[1ex]
        \text{s.t.} &~\eqref{OP0-APVCons1},~\eqref{OP0-APVCons2},
    \end{align}
\end{subequations}
where $\bm A = \sum\nolimits_{k=1}^K (\bm A_k + \tilde{\bm A}_k)$, $\bm v^i = \sum\nolimits_{k=1}^K (\bm v^i_k + 2 \tilde{\bm v}^i_k)$, and $C^i = \sum\nolimits_{k=1}^K (C^i_k - 2 \tilde C^i_k + 1)$. Since \eqref{OP0-APVCons1} and \eqref{OP0-APVCons2} are linear constraints and $\bm A$ is a positive semi-definite matrix, therefore \eqref{Opt-x3} is a convex optimization problem and can be efficiently solved using CVX \cite{CVX}.
\end{appendices}

\end{document}